\documentclass[superscriptaddress,twocolumn,amsmath,amssymb,showpacs,floatfix,prb]{revtex4}

\usepackage{graphicx}
\usepackage{dcolumn}
\usepackage{bm}
\usepackage{rotating}
\usepackage{booktabs}
\usepackage{color}
\usepackage{graphics}
\usepackage{array}
\usepackage{units}
\usepackage{multirow}
\usepackage[small,it]{caption}%small captions, in italics
\usepackage{fontenc}

\newcommand{\CFO}{CuFeO$_2$}

\newcommand{\tna}{$T_{N1}$}
\newcommand{\tnb}{$T_{N2}$}
\newcommand{\fe}{Fe$^{3+}$}
\newcommand{\Bpar}{$B^{\parallel}$}
\newcommand{\Bpara}{$B^{\parallel}_{c1}$}
\newcommand{\Bparaup}{$B^{\parallel}_{c1\uparrow}$}
\newcommand{\Bparadown}{$B^{\parallel}_{c1\downarrow}$}
\newcommand{\Bparb}{$B^{\parallel}_{c2}$}
\newcommand{\Bparbup}{$B^{\parallel}_{c2\uparrow}$}
\newcommand{\Bparbdown}{$B^{\parallel}_{c2\downarrow}$}
\newcommand{\Bparc}{$B^{\parallel}_{c3}$}
\newcommand{\Bparcup}{$B^{\parallel}_{c3\uparrow}$}
\newcommand{\Bparcdown}{$B^{\parallel}_{c3\downarrow}$}
\newcommand{\Bpard}{$B^{\parallel}_{c4}$}
\newcommand{\Bpare}{$B^{\parallel}_{c5}$}
\newcommand{\Bpareup}{$B^{\parallel}_{c5\uparrow}$}
\newcommand{\Bparedown}{$B^{\parallel}_{c5\downarrow}$}
\newcommand{\Bparsat}{$B^{\parallel}_{sat}$}
\newcommand{\Bperp}{$B^{\perp}$}
\newcommand{\Bperpa}{$B^{\perp}_{c1}$}
\newcommand{\Bperpaup}{$B^{\perp}_{c1\uparrow}$}
\newcommand{\Bperpadown}{$B^{\perp}_{c1\downarrow}$}
\newcommand{\Bperpb}{$B^{\perp}_{c2}$}
\newcommand{\Bperpc}{$B^{\perp}_{c3}$}
\newcommand{\Bperpcup}{$B^{\perp}_{c3\uparrow}$}
\newcommand{\Bperpcdown}{$B^{\perp}_{c3\downarrow}$}

\begin{document}

\title{Mapping the $\mbox{\boldmath{$B$}}$,$\mbox{\boldmath{$T$}}$ phase diagram of frustrated metamagnet \CFO.}

\author{T.T.A. Lummen}
\affiliation{Zernike Institute for Advanced Materials, University of
Groningen, Nijenborgh 4, 9747 AG Groningen, The Netherlands}

\author{C. Strohm}
\altaffiliation{Current address: European Synchrotron Radiation
Facility (ESRF) - P.O. Box 220, 38043 Grenoble, France}
\affiliation{Institut N\'{e}el, CNRS et Universit\'{e} Joseph
Fourier, BP 166, F-38042, Grenoble Cedex 9, France}

\author{H. Rakoto$^{\dag}$}%here and in acknowledgements, also: LNCMP LAB WILL NEED TWO REPRINTS AFTER PUBLICATION
\affiliation{Laboratoire National des Champs Magn\'{e}tiques
Intenses-Toulouse (LNCMI-Toulouse), 143 avenue de Rangueil, 31400
Toulouse, France}

%\author{A.A. Nugroho}
%\affiliation{Departemen Fisika FMIPA, Institut Teknologi Bandung,
%Jl. Ganesa 10, Bandung 40132, Indonesia}

%\author{I.P. Handayani}
%\affiliation{Zernike Institute for Advanced Materials, University of
%Groningen, Nijenborgh 4, 9747 AG Groningen, The Netherlands}

%\author{G. Dhalenne}
%\affiliation{Laboratoire de Physico-Chimie de l'Etat Solide, CNRS,
%UMR8182, Universit\'{e} Paris-Sud, B\^{a}timent 414, 91405 Orsay,
%France}

%\author{A. Revcolevschi}
%\affiliation{Laboratoire de Physico-Chimie de l'Etat Solide, CNRS,
%UMR8182, Universit\'{e} Paris-Sud, B\^{a}timent 414, 91405 Orsay,
%France}

\author{P.H.M. van Loosdrecht}
\affiliation{Zernike Institute for Advanced Materials, University of
Groningen, Nijenborgh 4, 9747 AG Groningen, The Netherlands}

\date{\today}

\begin{abstract}
The magnetic phase diagram of \CFO\ as a function of applied
magnetic field and temperature is thoroughly explored and expanded,
both for magnetic fields applied parallel and perpendicular to the
material's c-axis. Pulsed field magnetization measurements extend
the typical magnetic staircase of \CFO\ at various temperatures,
demonstrating the persistence of the recently discovered high field
metamagnetic transition up to \tnb\ $\approx 11$ K in both field
configurations. An extension of the previously introduced
phenomenological spin model used to describe the high field
magnetization process (\textit{Phys. Rev. B}, \textbf{80}, 012406
(2009)) is applied to each of the consecutive low-field commensurate
spin structures, yielding a semi-quantitative simulation and
intuitive description of the entire experimental magnetization
process in both relevant field directions with a single set of
parameters.
\end{abstract}


\maketitle

\section{Introduction}
\begin{figure*}[htb]
\centering
\includegraphics[width=12.5cm]{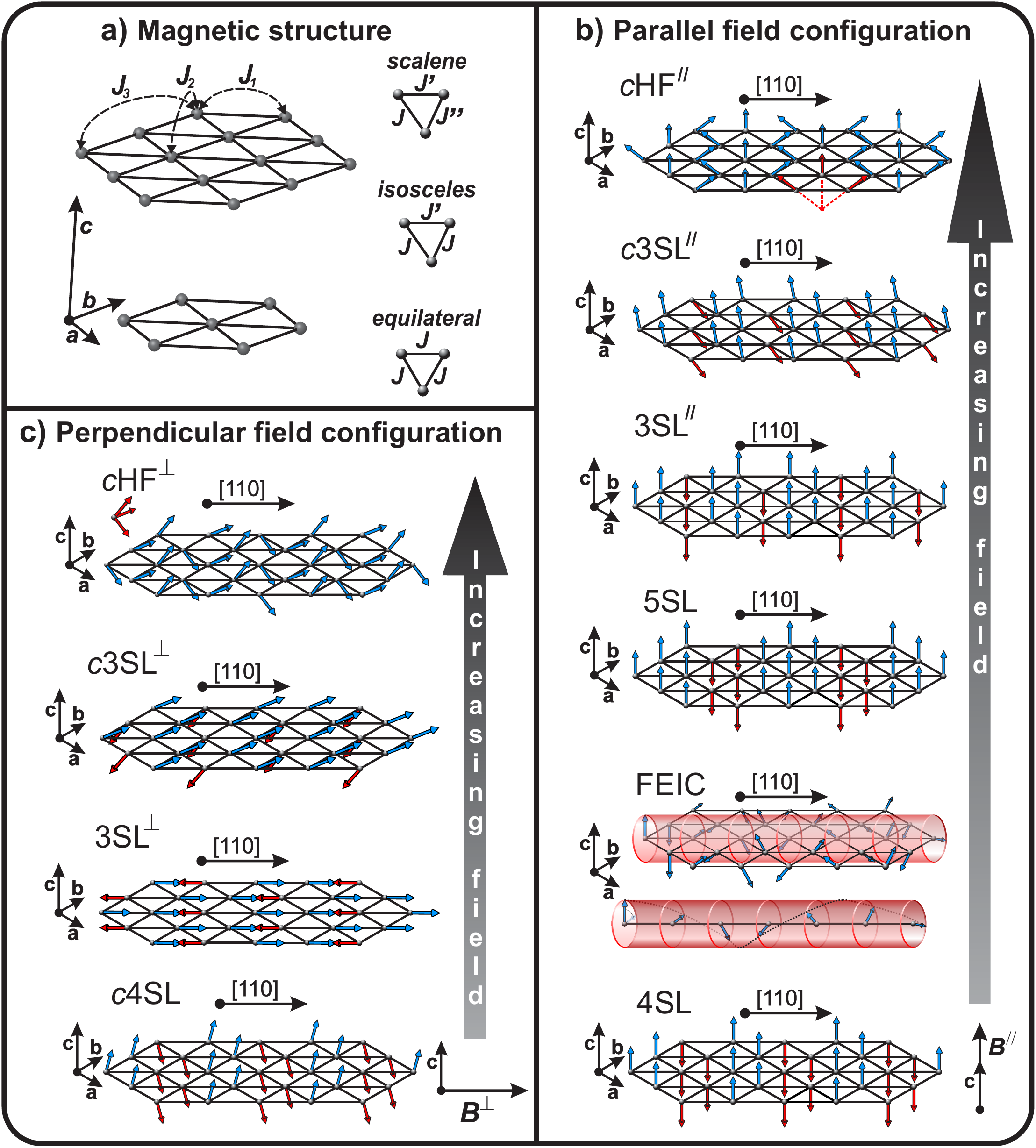}
\caption{\label{fig1} \small (Color online) \textbf{a)} Magnetic
structure of \CFO, space group \textit{R\={3}m}, $a=b=3.03$\AA,
$c=17.17$\AA. Only the magnetic Fe$^{3+}$ ions ($3d^5$, $S = 5/2$)
are depicted, illustrating the magnetic structure of quasi-separate
triangular layers. Different triangular symmetries as consecutively
occurring in \CFO\ are depicted on the right. \textbf{b)}
Successively adopted spin structures in the various phases of \CFO,
when subjected to an increasing applied magnetic field $B\parallel$
c. The higher field magnetic phases are proposed on the basis of a
recently reported classical spin model (PCS model, see text).
\textbf{c)} Analogous sequence of consecutively adopted spin
arrangements in \CFO\ for the $B\perp$ c configuration.}
\end{figure*}

One of the richest and most fascinating phenomena in magnetic
systems, geometrical frustration, occurs when the specific geometry
of an atomic lattice prevents, or frustrates, the simultaneous
minimization of all magnetic exchange interactions within the
system, thereby inducing a large magnetic degeneracy. With the
primary interactions of the magnetic system unable to select a
unique magnetic ground state, the magnetic behavior of frustrated
systems is dominated by secondary, often weaker interactions, which
can vary strongly even across closely related materials.
Consequently, the field of frustrated magnetism is characterized by
its vast richness and diversity, exotic magnetic states and low
temperature physics.\cite{ramirez94,schiffer96,diep94,greedan01} One
of the classic geometries in which this phenomenon readily manifests
itself is the triangular lattice with antiferromagnetic
interactions. In absence of significant secondary interactions,
classical spins on a triangular lattice antiferromagnet (TLA)
compromise in their 'desire' to align antiparallel and adopt a
noncollinear $120^{\circ}$ spin configuration at low temperatures,
an underconstrained, highly degenerate ground state.\cite{kadowaki90,collins97} The situation
can be quite different, however, in systems where secondary
interactions are significant, such as in the stacked delafossite
material \CFO, which consists of hexagonal Fe$^{3+}$, O$^{2-}$ and
Cu$^{+}$ layers (space group \textit{R\={3}m}, $a=b=3.03$\AA,
$c=17.17$\AA). As the Fe$^{3+}$ ($3d^5$, $S = 5/2$) ions are the
system's only magnetic constituents (Cu$^{+}$ and O$^{2-}$ have
filled electronic shells), and their spins interact
antiferromagnetically, the magnetic system corresponds to an
archetypical TLA at room temperature (Figure \ref{fig1}a).
Strikingly though, in contrast to other delafossite TLAs like
LiCrO$_{2}$, AgCrO$_{2}$ and CuCrO$_{2}$,\cite{kadowaki90,collins97}
\CFO\ adopts a collinear ground state at low temperatures.
\indent Based on the electronic configuration of the \fe\ ion
($^{6}S_{5/2}$, $L = 0$), the antiferromagnetic exchange
interactions within the system are expected to be isotropic, thus
yielding a pure Heisenberg TLA. The presence of a substantial
spin-lattice coupling in \CFO\ (the secondary interaction) however,
induces a low-temperature structural distortion through the 'spin Jahn-Teller'
effect\cite{yamashita00,tchernyshyov02,wang08}, hereby reducing the
spin state degeneracy in the system.
%This spin-lattice coupling has its origin in the distance dependence of the magnetic exchange
%coupling between two sites $i$ and $j$,
%$J(r_{ij})\mbox{\boldmath{$S$}}_{i}\cdot \mbox{\boldmath{$S$}}_{j}$,
%and is also known as magnetoelastic exchange. The symmetry lowering
%effect of this coupling is analogous to that of the spin-Peierls
%reminiscent of the collective Jahn-Teller effect in crystalline
%solids. Analogous to the situation in the traditional Jahn-Teller
%case, where the electronic ground state is degenerate, the high
%symmetry of the system induces a degeneracy in the spin states here.
%In accordance with the original Jahn-Teller theorem, the system can
%thus lower its ground state energy by undergoing a structural
%distortion. Because in this case the spin variables induce the
%degeneracy and drive the distortion, it is termed the spin
%Jahn-Teller effect.
The structural symmetry of the \CFO\ lattice
is first lowered from the hexagonal \textit{R\={3}m} space group to the
monoclinic \textit{C2/m} space group at \tna\ $\approx 14$ K, to be
further reduced to a lower monoclinic symmetry at \tnb\ $\approx 11$
K.\cite{terada06_1,ye06,terada06_2,terada07_1} Magnetically, \CFO\
undergoes a transition from its paramagnetic (PM) phase to a
partially disordered, incommensurate (PDIC) magnetic phase at \tna\,
where a sinusoidally amplitude-modulated magnetic structure with a
temperature dependent propagation wave vector ($q$ $q$ $0$) is
adopted.\cite{mekata93,mitsuda98} Another magnetic phase transition
at \tnb\ brings the system into its collinear four-sublattice (4SL)
ground state, in which the spins align (anti-)parallel to the
\textit{c}-axis, adopting an in-plane two-up two-down order, as illustrated in Figure
\ref{fig1}b.\cite{mitsuda91,mekata92} To avoid confusion, we will
refer to crystallographic directions using the hexagonal description
depicted in Figure \ref{fig1} throughout the paper.\newline
\indent The stabilization of the collinear 4SL state in \CFO\ proved
to be one of its most puzzling issues. Initially, the system was
described as a two-dimensional (2D) Ising TLA with exchange
interactions up to the third nearest-neighbors. The first ($J_{1}$),
second ($J_{2}$) and third ($J_{3}$) in-plane nearest-neighbor
interactions were estimated to compare as $J_{2}/J_{1} \approx 0.5$
and $J_{3}/J_{1} \approx 0.75$ in this
model\cite{mekata92,ajiro94,takagi95,mitsuda98,mitsuda99}, with
$J_{1}$ corresponding to approximately 1.2
meV.\cite{ye07,petrenko05,ajiro94} There is, however, \textit{a
priori} no physical justification for the assumed Ising nature of
the magnetic moments. Such an assumption is also inconsistent with
magnetic susceptibility measurements, which show highly isotropic
behavior above \tna\ in
\CFO.\cite{ajiro94,zhao96,petrenko05,kimura06} Nonetheless, the
magnetic properties below \tna\ are unmistakably strongly
anisotropic. The recent discovery of the low temperature structural
distortion offers an alternate picture, as it results in a lattice
of scalene triangles in the basal plane (see Fig. \ref{fig1}a),
which splits the first nearest-neighbor interaction within every
triangle into three unequal exchange interactions, lowering the
energy of the 4SL
state.\cite{terada06_1,ye06,terada06_2,terada07_1}. Perhaps more
importantly, the distortion has been argued induce a small easy axis
anisotropy along the c-axis as well, further stabilizing the
collinear ground state.\cite{terada07_3} Experimentally, a small
single-ion anisotropy interaction was estimated by fitting a 3D
Heisenberg Hamiltonian with a single-ion anisotropy term to the
spin-wave dispersion along the $c$ axis below \tna, which supports
the picture of distortion-induced anisotropy.\cite{ye07,fishman08}
As will be confirmed below, the combination of this weak magnetic
anisotropy and the relatively strong spin-phonon coupling in \CFO\
can explain its observed Ising-like
behavior.\cite{wang08,lummen09}\newline
\indent Arguably the most fascinating physical properties arise when
\CFO\ is subjected to an external magnetic field below \tnb. Upon
increasing applied magnetic field along the c axis ($B\parallel$ c),
the material has been shown to undergo a series of magnetic
transitions at \Bpara\ $\simeq$ 7 T, \Bparb\ $\simeq$ 13 T, \Bparc\
$\simeq$ 20 T, \Bpard\ $\simeq$ 34 T and \Bpare\ $\simeq$ 53 T,
before ultimately reaching saturation around \Bparsat\ $\simeq$ 70
T.\cite{ajiro94,petrenko00,mitsuda00_1,petrenko05,terada06_3,kimura06,terada07_3,lummen09,quirion09}
Corresponding magnetic structures between the successive transitions
(see Fig. \ref{fig1}b) were shown to be a proper helical magnetic order
with an incommensurate in-plane wave vector, which also carries a ferroelectric
moment\cite{kimura06,arima07,nakajima07,nakajima08} (\Bpara\ \textless\ \Bpar\ \textless\ \Bparb, FEIC), and a
collinear five-sublattice (5SL) phase where the spins again align
(anti-)parallel to the c axis, adopting a three-up two-down order
(\Bparb\ \textless\ \Bpar\ \textless\ \Bparc).\cite{mitsuda00_2,petrenko00} Spin structures at higher
fields have not yet been experimentally determined due to the
demanding experimental requirements. In a recent work, we have
reported pulsed field magnetization measurements, revealing the
retrieval of virtually isotropic magnetic behavior above an additional
phase transition at \Bpare.\cite{lummen09} A corresponding anomaly was
subsequently observed at somewhat lower fields in pulsed-field ultrasonic velocity
measurements by Quirion \emph{et al.}\cite{quirion09}, confirming its proposed magneto-elastic nature.
On the basis of a phenomenological classical
spin model (PCS), the spin structures in the high field magnetic
phases were suggested to correspond to a collinear three-sublattice
(3SL$^{\parallel}$, \Bparc\ \textless\ \Bpar\ \textless\ \Bpard), an anisotropic \textit{canted} three-sublattice
(\textit{c}3SL$^{\parallel}$, \Bpard\ \textless\ \Bpar\ \textless\
\Bpare), and an isotropic canted high-field magnetic order
(\textit{c}HF$^{\parallel}$, \Bpar\ \textgreater\ \Bpare), as
depicted in Figure \ref{fig1}b.\cite{lummen09}
\newline
\indent Illustrating the low temperature anisotropy in the material,
the magnetism in \CFO\ evolves quite differently when it is
subjected to a magnetic field perpendicular to the c axis ($B\perp$
c) below \tnb, showing only two transitions up to 40 T, at \Bperpa\
$\simeq$ 24 T and \Bperpb\ $\simeq$ 30
T.\cite{ajiro94,petrenko00,petrenko05,terada07_3} Our recent results
also revealed a high field magnetic transition for this field
configuration, at \Bperpc\ $\simeq$ 51.6 T.\cite{lummen09} Apart
from the low field 4SL structure, the corresponding magnetic
structures have not yet been experimentally determined. Based on the
magnetization measurements and the aforementioned PCS model, the
magnetic structure has been proposed to undergo consecutive spin
rearrangements from a \textit{canted} 4SL order
(\textit{c}4SL$^{\perp}$, with spins tilted away from the
$c$-direction) to a collinear 3SL phase (3SL$^{\perp}$, with spins
in the basal plane) at \Bperpa, to a \textit{canted} 3SL order at
\Bperpb\ (\textit{c}3SL$^{\perp}$), and finally to the isotropic
canted high field configuration (\textit{c}HF$^{\perp}$) at \Bperpc\
(see Fig. \ref{fig1}c).\newline
\indent As is clear from above disquisition, the magnetic behavior
of \CFO\ as a function of temperature and applied magnetic field has
proven very rich and has yielded unanticipated, fascinating new
insights. Following our recent results, this work aims to thoroughly
map out and extend the intricate $B$,$T$ phase diagrams of \CFO\ up
to 58 T and \tnb\ $\approx 11$ K, for both for the $B\parallel$ c
and the $B\perp$ c configuration. Furthermore, by applying the
recently introduced PCS model to all commensurate sublattice phases
occurring in \CFO, an adequate description of the entire experimental
magnetization process in both field configurations and an intuitive understanding of the magnetic
behavior in \CFO\ is provided.

\section{Experimental}
\subsection{Sample preparation}
A high quality, single crystalline rod of \CFO\ was synthesized
using the floating zone technique, following the procedure described
by Zhao \textit{et al.}\cite{zhao96} A $^{57}$Fe-enriched starting
material ($^{57}$Fe$_{2}$O$_{3}$, $^{57}$Fe \textgreater\ 95.5\%)
was used in the synthesis, to facilitate nuclear forward scattering
experiments described elsewhere.\cite{strohm10_2} X-ray Laue
diffraction was employed to orient the \CFO\ single
crystal. Next, small cuboid samples (5 x 1 x 1 mm$^{3}$), with
long sides oriented parallel (35.9 mg) and perpendicular (42.1 mg)
to the crystallographic c-axis, respectively, were prepared from the
single crystal. Further characterization, including $^{57}$Fe
M\"{o}ssbauer spectroscopy, Raman spectroscopy and SQUID
magnetometry, also yielded experimental data in excellent agreement
with literature on \CFO, confirming the high sample quality. The
same samples were used in all measurements reported here and in a
previous work.\cite{lummen09}

\subsection{Instrumentation}
High (pulsed) magnetic field magnetization measurements, up to a
maximum field of 58.3 T were performed at the 'Laboratoire National
des Champs Magn\'{e}tiques Puls\'{e}s' in Toulouse, France. The
obtained magnetization data were accurately scaled through a least
squares fit to low field measurements (up to 10 T), performed on a
well calibrated static (dc) magnetic field setup (using the
extraction technique) of the 'Institut N\'{e}el' in Grenoble,
France. The accuracy in the scaling procedure was such that it
introduces an uncertainty of $\pm$ 0.3\% in all magnetization values
determined from the pulsed field experiments. The temperature dependence
of dc magnetic susceptibilities of oriented single crystalline \CFO\ cuboids
was measured in various constant magnetic fields (up to 7 T) using a
Quantum Design MPMS magnetometer.
\section{Results and Discussion}
\subsection{Magnetization in pulsed magnetic fields}
\subsubsection{Parallel field configuration ($B\parallel$ c)}

\begin{figure*}[htb]
\centering
\includegraphics[width=12.5cm]{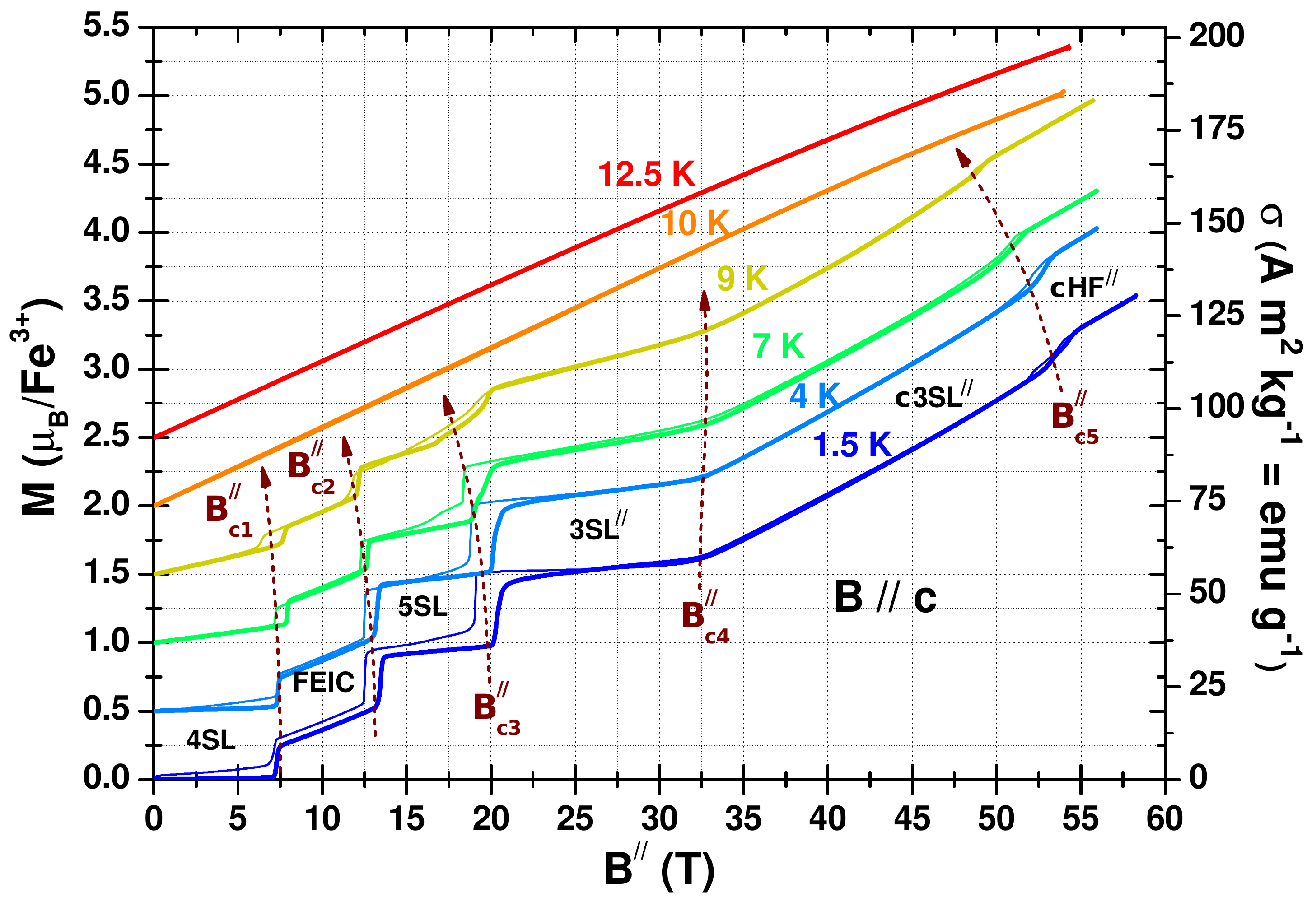}
\caption{\label{fig2} \small (Color online) Magnetization
measurements in pulsed magnetic fields at various temperatures. The
magnetic field is applied in the direction parallel to the c-axis.
The various curves are offset by consecutive multiples of 0.5
$\mu_{B}$/Fe$^{3+}$ with increasing temperature for clarity. Thick
and thin lines represent sample magnetization in increasing and
subsequently decreasing magnetic field, respectively. Dashed arrows
indicate the temperature dependence of the various magnetic
transitions (See also Fig. \ref{fig3}).}
\end{figure*}

Figure \ref{fig2} depicts the magnetization curves up to 58.3 T for
various temperatures below \tna, where the applied magnetic field
$B$ is parallel to the c-axis (\Bpar). As \Bpar\ increases, several
successive metamagnetic steps are observed, in excellent agreement
with previously reported
results.\cite{ajiro94,petrenko00,mitsuda00_1,petrenko05,terada06_3,kimura06,terada07_3,lummen09,quirion09}
At 1.5 K, the system shows magnetic phase transitions at \Bpara\
$\simeq$ 7.2 T (4SL to FEIC phase transition), \Bparb\ $\simeq$ 13.0
T (FEIC $\rightarrow$ 5SL), \Bparc\ $\simeq$ 19.7 T (5SL
$\rightarrow$ 3SL$^{\parallel}$), \Bpard\ $\simeq$ 32.4 T
(3SL$^{\parallel}$ $\rightarrow$ \textit{c}3SL$^{\parallel}$) and
\Bpare\ $\simeq$ 53.3 T (\textit{c}3SL$^{\parallel}$ $\rightarrow$
\textit{c}HF$^{\parallel}$). The three transitions at lowest
critical fields \Bpara, \Bparb\ and \Bparc\ are all accompanied by
large magnetization steps and exhibit significant hysteresis
(\Bparaup\ = 7.27 T, \Bparadown\ = 7.15 T, \Bparbup\ = 13.44 T,
\Bparbdown\ = 12.51, \Bparcup\ = 20.32 T and \Bparcdown\ = 19.08 T
at 1.5 K), indicating their first order nature. In contrast, at the
fourth magnetic transition (\Bpard\ $\simeq$ 32.4 T), the
$M$(\Bpar)-curve shows only a change in slope, suggesting this
transition is of second order, which is consistent with synchrotron
x-ray diffraction results.\cite{terada06_3,terada07_3} The high
field transition at \Bpare\ is again of first order nature, as
illustrated by its hysteresis: \Bpareup\ = 53.78 T and \Bparedown\ =
52.88 T at 1.5 K. The existence of this high field transition was
recently confirmed in ultrasonic velocity measurements \cite{quirion09} and can
also be seen in previous magnetization data recorded by Ajiro
\textit{et al.}, who measured the magnetization of a powder sample
of \CFO\ at 8 K in a single turn coil measurement up to 100
T.\cite{ajiro94} Though it is obscured in their $M$,$B$-curve, presumably due to
the polycrystalline nature of the sample, a clear feature can be
seen around $\sim$ 52 T in the corresponding ($dM/dB$) vs. $B$
graph.\newline
\indent In the 4SL phase, the magnetization is close to zero, as
expected for the two-up two-down structure (Fig. \ref{fig1}b, 4SL).
In the FEIC phase, $M$ increases linearly with \Bpar\ as observed
before\cite{terada04_1,petrenko05,kimura06,terada07_3,lummen09},
signaling a continuous reorientation of the spin system in the
spiral phase (Fig. \ref{fig1}b, FEIC). In the 5SL phase, $M$ is
almost constant, at a value approximately equal to one-fifth of the
5 $\mu_{B}$/Fe$^{3+}$ saturation value, in good agreement with the
three-up two-down structure (Fig. \ref{fig1}b, 5SL). Between \Bparc\
and \Bpard, $M$ is again almost independent of \Bpar, having a value
close to 1/3rd of the saturation-value, while between \Bpard\ and
\Bpare\ the magnetization again increases linearly with \Bpar,
indicating another continuous reorientation of the spin system.
Based on these observations and the PCS model, these phases have
been proposed to correspond to a collinear three-sublattice (Fig.
\ref{fig1}b, 3SL$^{\parallel}$, two-up one down) and a
\textit{canted} three-sublattice phase (Fig. \ref{fig1}b,
$\textit{c}$3SL$^{\parallel}$), respectively.\cite{terada06_3,
terada07_3,lummen09} At \Bpare, the system undergoes another first
order transition, where the magnetization exhibits an abrupt jump.
Above \Bpare, the magnetization shows a steady linear increase up to
the highest field measured, 58.27 T. At this point $M$ has taken a
value of 3.54 $\mu_{B}$/Fe$^{3+}$ (at 1.5 K), close to the $\simeq$
3.7 $\mu_{B}$/Fe$^{3+}$ value for the powder sample measured at 8 K
by Ajiro \textit{et al.}\cite{ajiro94} As the system has regained
isotropic behavior above this transition, the spin structure in this
\textit{c}HF$^{\parallel}$ phase has been proposed to be isotropic,
such as e.g. the canted 120\ensuremath{^\circ} configuration
depicted in Fig. \ref{fig1}b, where the projection of the spins in
the basal plane retains the typical 120\ensuremath{^\circ} configuration
while the out of plane spin-components grow with
B$^{\parallel}$. In their recent paper, Quirion \emph{et al.} proposed
a similar, though slightly incommensurate 120\ensuremath{^\circ}-like
spin structure based on Landau free energy considerations.\cite{quirion09}
\newline
\begin{figure*}[htb]
\centering
\includegraphics[width=10cm]{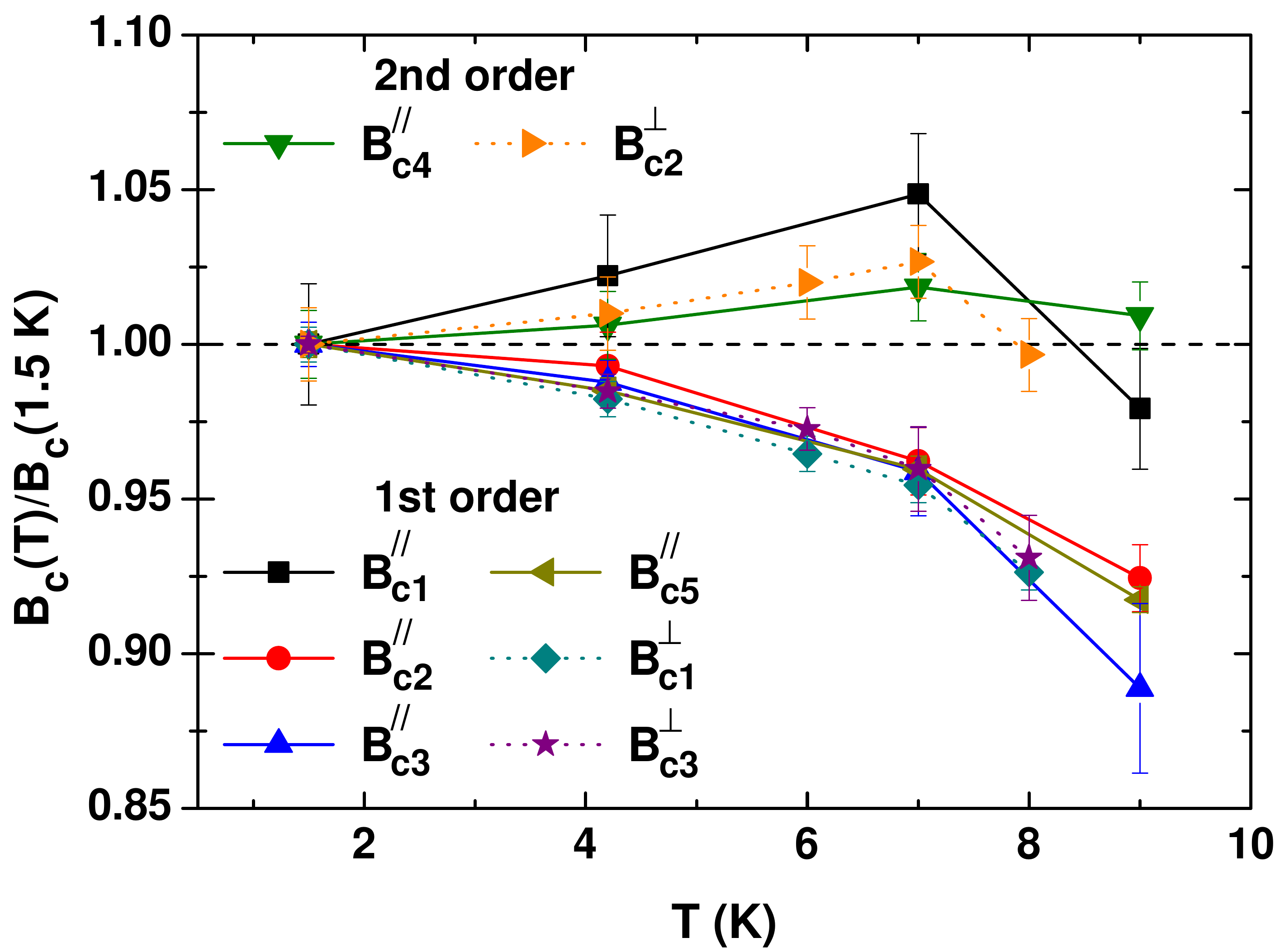}
\caption{\label{fig3} \small (Color online) Temperature dependence
of the critical fields corresponding to the various magnetic
transitions, for both configurations ($B\parallel$ c and $B\perp$
c). Critical field values (midpoints of hysteresis loop for first
order transitions) are normalized by their corresponding 1.5 K
values.}
\end{figure*}
\indent As the temperature increases, the general features of the
$M$,$B$-curve survive, though magnetic steps are broadened over
an increasingly wide field range, hysteresis widths are reduced and
plateau phases acquire increasing slopes. As the temperature
approaches \tnb, the characteristic staircase features of the
magnetization smooth out and $M$ increases (quasi-)linearly with
$B$, deviating from this behavior only at high magnetic fields,
close to saturation. The fact that this appears to occur already
just below \tnb\ is ascribed to a slight offset of the corresponding
temperature sensor at these temperatures, as transition temperatures
measured in susceptibility experiments on the same sample (see
below) are in accordance with literature values. A striking feature
is the temperature dependence of the various magnetic transitions
(indicated by the dashed arrows in Figure \ref{fig2}). Figure
\ref{fig3} shows the relative variation of the corresponding
critical magnetic fields with temperature. With the exception of the
lowest field-induced transition, all (first order) transitions
exhibiting hysteresis show identical behavior; a continuous decrease
of the corresponding critical field (\Bparb, \Bparc\ and \Bpare,
respectively) with increasing temperature. In contrast, the critical
field of the second order transition (\Bpard) proves rather
temperature independent, once more indicating its different nature.
\subsubsection{Perpendicular field configuration ($B\perp$ c)}
\begin{figure*}[htb]
\centering
\includegraphics[width=12.5cm]{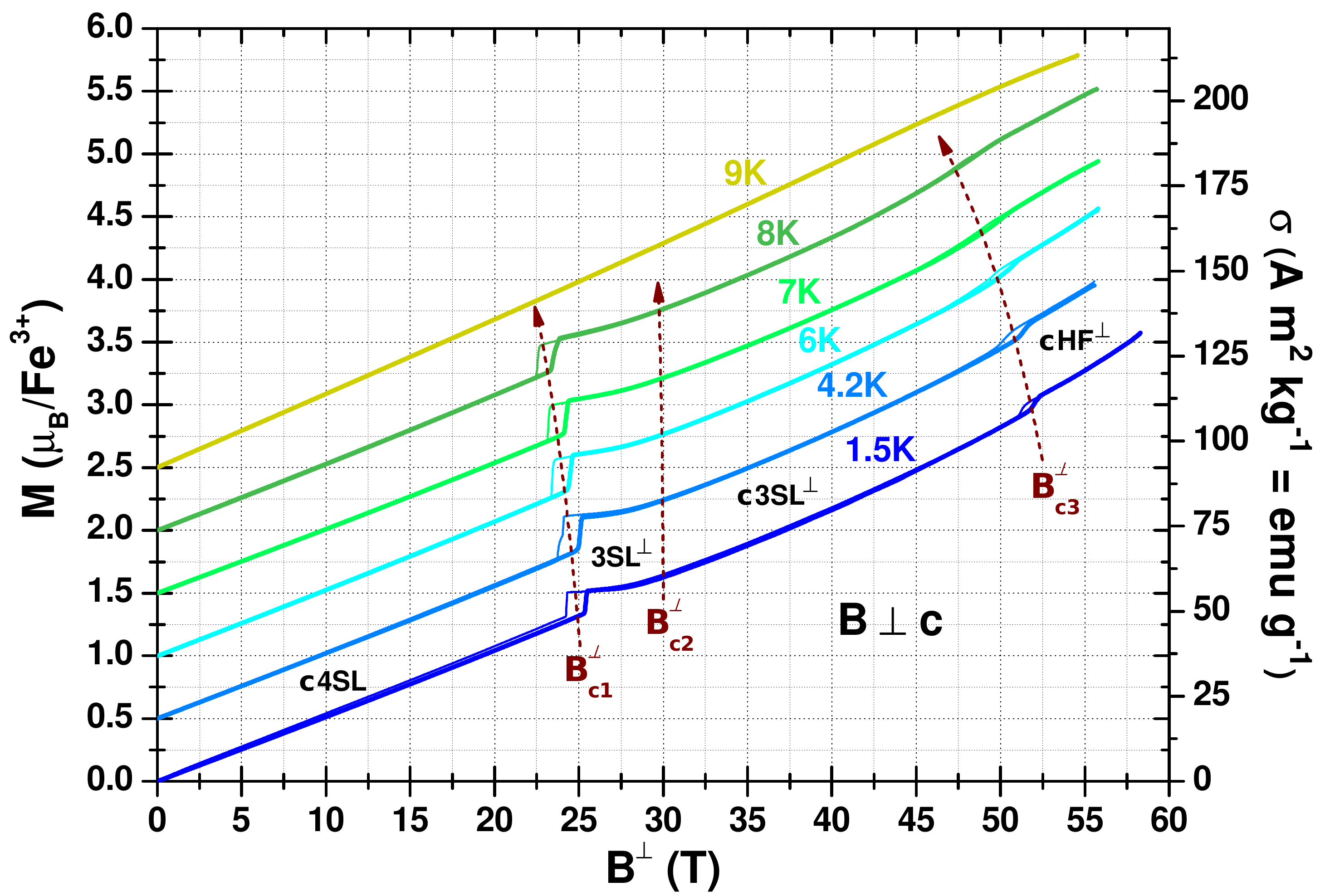}
\caption{\label{fig4} \small (Color online) Magnetization
measurements in pulsed magnetic fields at various temperatures.
Here, the applied magnetic field is perpendicular to the c direction
of the crystal. The various curves are offset by consecutive
multiples of 0.5 $\mu_{B}$/Fe$^{3+}$ with increasing temperature for
clarity. Thick and thin lines represent sample magnetization in
increasing and subsequently decreasing magnetic field, respectively.
Dashed arrows indicate the temperature dependence of the various
metamagnetic transitions.}
\end{figure*}

Figure \ref{fig4} shows the magnetization process up to 58.3 T for
various temperatures below \tna, for the perpendicular configuration
($B$ $\perp$ $c$). As for the parallel configuration, the
magnetization curves are in excellent agreement with earlier
observations.\cite{ajiro94,petrenko00,petrenko05,terada07_3,lummen09}
With increasing \Bperp, the magnetization shows a steady linear
increase up to \Bperpa\ ($\simeq$ 24.8 T at 1.5 K), suggesting a
slight continuous canting of the 4SL spins from the $c$ direction,
toward the basal ($a$,$b$) plane (\textit{c}4SL, Fig. \ref{fig1}c).
Indeed, neutron diffraction data have confirmed the stability of
this \textit{c}4SL magnetic structure up to at least 14.5
T.\cite{petrenko00} At \Bperpa, the system undergoes a first order
magnetic transition to a plateau state, which shows significant
hysteresis (at 1.5 K, \Bperpaup\ = 25.40 T and \Bperpadown\ = 24.27
T). The magnetization in this plateau state is rather independent of
\Bperp\ at an average value of $\simeq$ 1.53 $\mu_{B}$/Fe$^{3+}$,
close to 1/3rd of the saturation value, implying a three-sublattice
state with spins in the basal plane, directed along \Bperp\ (Fig.
\ref{fig1}c, 3SL$^{\perp}$). This spin configuration was recently
confirmed using numerical minimization of the PCS
model.\cite{lummen09} After undergoing a second order phase
transition at \Bperpb\ $\simeq$ 30.0 T (at 1.5 K), $M$ once again
increases (quasi-)linearly with \Bperp, which in turn implies a
continuous reorientation of the moments away from collinearity. Due
to the nonzero easy axis anisotropy at these fields, the slope in
this canted 3SL phase (Fig. \ref{fig1}c, \textit{c}3SL$^{\perp}$)
differs from that in the same field interval for the parallel
configuration. At \Bperpc\ $\simeq$ 51.6 T (1.5 K), another
magnetic transition is observed, similar to that at \Bpare\ in
the parallel configuration. As in that configuration, the high field
transition here consists of a first order metamagnetic step, which
exhibits hysteresis (at 1.5 K, \Bperpcup\ = 52.02 T and \Bperpcdown\
= 51.18 T). At 1.5 K, $M$ jumps to $\simeq$ 3.1 $\mu_{B}$/Fe$^{3+}$
at \Bperpcup, after which it resumes a steady increase, in line with
a noncollinear spin arrangement. In fact, based on the absence of
anisotropy in this canted HF phase (\textit{c}HF$^{\perp}$), the PCS
model predicts a spin structure analogous to that for $B\parallel$
c, as sketched in Fig. \ref{fig1}c. The fact that the additional
transition occurs at slightly lower critical field in the
perpendicular configuration (\Bperpc\ $\simeq$ 51.6 T vs. \Bpare\
$\simeq$ 53.3 T) explains the broadness of the feature around $\sim$
52 T in the aforementioned ($dM/dB$),$B$-curve of the polycrystalline sample
of Ajiro \textit{et al.}\cite{ajiro94}\newline
\indent With increasing temperature, the general features of the
$M$,$B$-curve remain intact, although the plateau phase acquires an
increasing slope. Furthermore, as for the parallel case, the
transition features are smoothed out upon approaching \tnb, and
deviation from this behavior only occurs upon approaching
saturation. Again, the apparent small temperature mismatch with
respect to susceptibility measurements (below) is attributed to a
slight offset of the temperature sensor at temperatures close to
\tnb. Also for $B\perp$ c, the temperature dependence of the various
critical fields correlates to the nature of the corresponding
transitions (See Fig. \ref{fig3}); first order transitions (at
\Bperpa\ and \Bperpc) exhibit the same relative decrease with
temperature as \Bparb, \Bparc\ and \Bpare, while the second order
transition (at \Bperpb) shows a much weaker temperature dependence.

\subsubsection{Progressive symmetry increase}

The strong coupling between spin and lattice degrees of freedom is a
key ingredient in the description of the magnetization process of
\CFO. Recently, Terada \textit{et al.}\cite{terada06_3,terada07_3}
showed the strong correlation between the lattice parameters and the
magnetization in applied field in both configurations. For
$B\parallel$ c, coinciding with the metamagnetic steps at \Bpara,
\Bparb, and \Bparc, the lattice undergoes corresponding
discontinuous contractions along the [110] direction, while changes
in the [$\bar{1}10$] direction are much smaller. In addition, the
lattice has been shown to increase its symmetry at \Bparb, where the
scalene triangle distortion is partially relieved, resulting in a
lattice of isosceles triangles (Fig. \ref{fig1}a).\cite{terada06_2}
The lattice parameter along [110] mirrors the behavior of the
magnetization in applied field; within the collinear phases it
remains practically constant and in the noncollinear phases the
lattice continuously contracts with increasing field (and
magnetization). These striking observations can be rationalized as
follows: in zero field, the spin-lattice coupling induces the
scalene triangle distortion and a magnetic easy axis along the $c$
direction, thereby reducing the magnetic energy at the expense of
elastic energy. As \Bpar\ increases however, a growing tendency for
parallel spin alignment in the field direction develops, thereby
successively reducing the degree of magnetic frustration (the driving
force for the distortion). Thus, as the gain in magnetic exchange energy
is successively reduced with \Bpar, the system rebalances the magnetic and
elastic energies associated with the lattice distortion along with every
spin rearrangement. As a result, the system exhibits a progressive lattice
contraction along [110], which mirrors the changes in magnetization.\newline
\indent Since the induced magnetic anisotropy in the material is
also directly coupled to the lattice distortion, one may expect the
strength of the induced single-ion anisotropy to diminish
accordingly with $M$, undergoing steps across first order
transitions and continuously decreasing in (\textit{quasi-})linear
phases. Indeed, as shown in Figure 2 of our recent
paper\cite{lummen09}, which shows the $M$,$B$-curves for both the
parallel and perpendicular configuration at 1.5 K, the system's
response to an applied field becomes more and more isotropic as $B$
increases. Moreover, above both high field transitions, recently
confirmed to be magneto-elastic in nature, the system was even
found to behave almost completely isotropic, consistent with a vanishing
easy axis anisotropy and the retrieval of an undistorted equilateral
triangular lattice.
\subsection{Magnetic susceptibility in constant fields}
\begin{figure*}[htb]
\centering
\includegraphics[width=15.0cm]{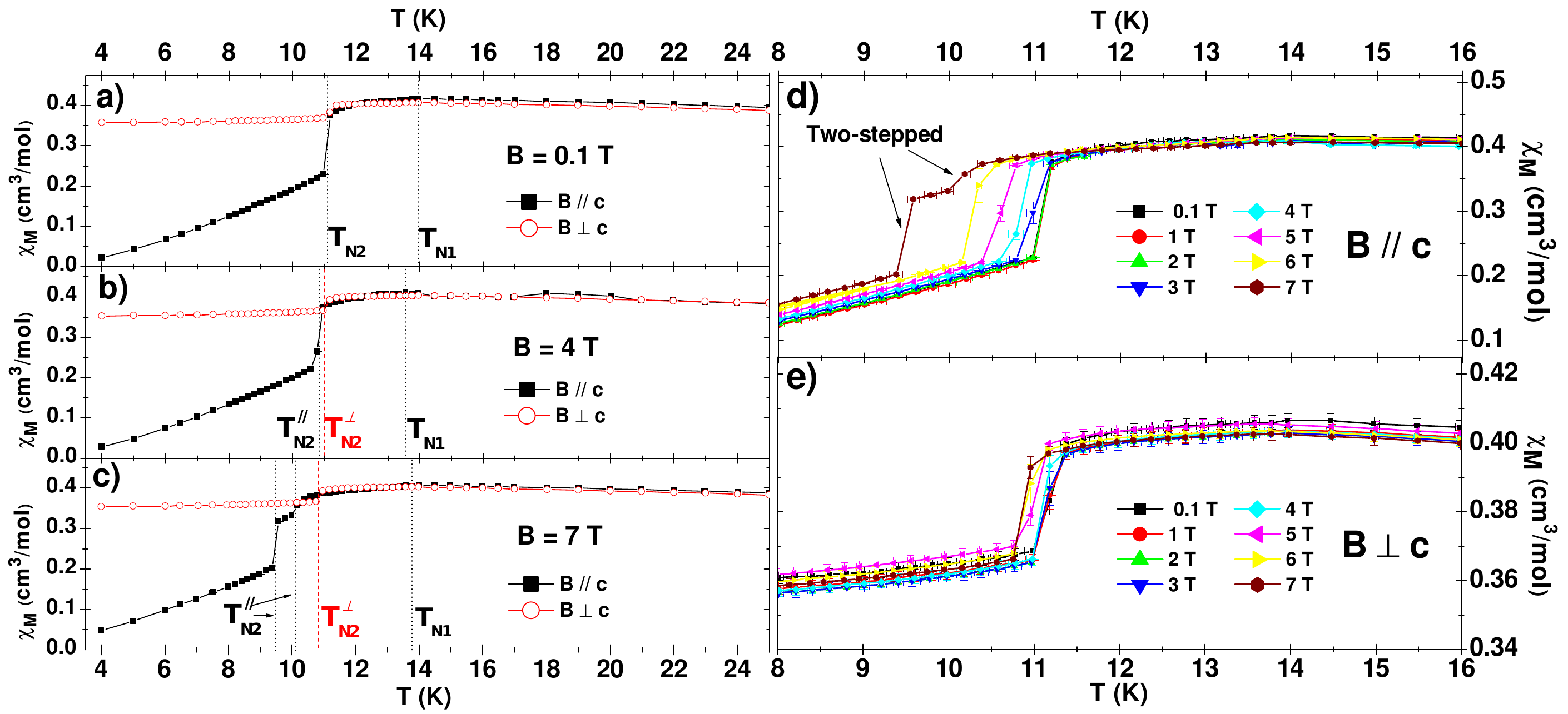}
\caption{\label{fig5} \small (Color online) Magnetic susceptibility
versus temperature in various constant magnetic fields for both the
parallel and the perpendicular configuration. Panels \textbf{a)-c)}
compare the magnetic susceptibility for $B\parallel$ c and $B\perp$
c as a function of temperature at selected fields up to 7 T.
Vertical dotted/dashed lines indicate magnetic transition
temperatures, as deducted from corresponding $d\chi_{M}$/$dT$ data.
The zoom-ins on the data in panels \textbf{i)} and \textbf{j)}
depict the field dependence of the magnetic susceptibility in
parallel and perpendicular configurations, respectively.}
\end{figure*}
In order to supplement the magnetic phase diagrams of \CFO\ and to
further elucidate its magnetic behavior, the temperature dependence
of dc magnetic susceptibilities in various constant magnetic fields
was measured in both field configurations. Panels
\textit{a})-\textit{c}) of Figure \ref{fig5} compare the low
temperature magnetic susceptibility curves for the two field
orientations in applied fields of 0.01, 4 and 7 T, respectively.
Consistent with previous measurements, both $\chi_{M}^{\parallel}$
($B\parallel$ c) and $\chi_{M}^{\perp}$ ($B\perp$ c) show a diffuse
maximum at \tna\ $\simeq$ 13.5 K and a subsequent abrupt drop at
\tnb\ $\simeq$ 11.2 K upon decreasing temperature.\cite{ajiro94,zhao96,terada04_2,petrenko05}
Above \tnb\ the susceptibility is isotropic, for all applied fields measured. As
expected for an ordered antiferromagnet, $\chi_{M}^{\parallel}$
approaches zero with decreasing temperature below \tnb, while
$\chi_{M}^{\perp}$ remains almost constant after the initial drop at
\tnb. The field dependence of the magnetic susceptibility,
visualized in panels \textit{d}) and \textit{e}) for the parallel
and perpendicular configuration, respectively, shows the invariance
of \tna\ with applied field for both configurations. Though
relatively field independent for the perpendicular configuration,
\tnb\ shifts to lower temperatures as the applied magnetic field
approaches \Bpara\ ($\simeq$ 7.2 T) in the parallel case. This
difference can be regarded as a consequence of the lower
susceptibility in the ordered phase for $B\parallel$ c, which is
unfavorable toward the Zeeman interaction, which becomes
increasingly strong with $B$. Thus, with increasing $B^{\parallel}$
the magnetic ordering transition at $T^{\parallel}_{\text{N}2}$ is
shifted to lower temperature. For the perpendicular case, the
susceptibility drop across $T^{\perp}_{\text{N}2}$ is only marginal,
ergo the corresponding temperature down-shift is far less
pronounced.
\newline \indent As is clear from panels \textit{d}) and \textit{e)},
the transition at $T^{\parallel}_{\text{N}2}$ and 7 T acquires a
double feature, indicating the process becomes two-stepped. This
points toward the presence of an intermediate phase between the two
steps. Based on the constructed phase diagram presented below
(Figure \ref{fig6}), this intermediate phase is identified as the
helical FEIC phase, as the phase boundaries of both the 4SL and FEIC
phases bend toward the $T^{\parallel}_{\text{N}2}(B)$ line at these
temperatures.

\subsection{Phase diagrams}

With the phase transition data obtained above in hand, the
experimental phase diagram of \CFO\ as a function of applied field
and temperature can be assembled for both field configurations.
Magnetic transition fields and temperatures are defined through the
position (center) of corresponding anomalies in the derivatives of
the pulsed field magnetization ($\partial$$M$/$\partial$$B$) and
susceptibility ($\partial$$\chi_{M}$/$\partial$$T$) curves,
respectively.
\begin{figure*}[htb]
\centering
\includegraphics[width=16.0cm]{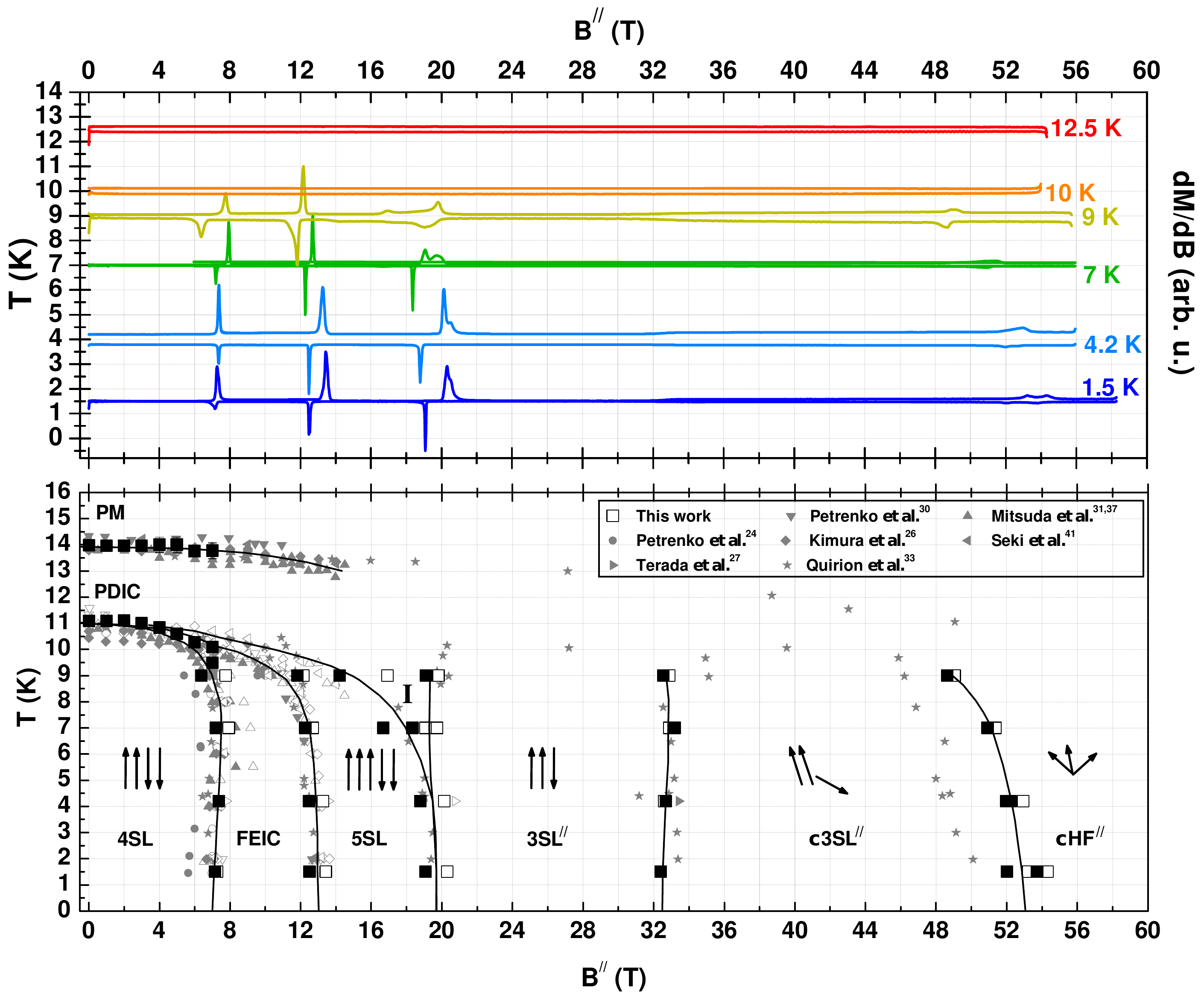}
\caption{\label{fig6} \small (Color online) Top panel: Differential
magnetization curves for $B\parallel$ c. Colored solid and dotted
lines depict (normalized) $dM$/$dB$-curves as measured in increasing
and subsequently decreasing magnetic field, respectively.
$dM$/$dB$-curves measured in decreasing magnetic field are inverted
for clarity. Additionally, each curve has an offset equal to its
corresponding temperature . Lower panel: $B$,$T$ phase diagram of
\CFO\ for the case where $B$ is parallel to the c-axis. Large, black
squares depict magnetic transitions as observed in this work and
smaller gray symbols indicate previously reported transitions. Open
and closed symbols represent transitions observed in increasing and
decreasing ($B$ or $T$) ramps, respectively. Solid lines are
correspond to proposed phase boundaries. Region $I$ corresponds to
an observed intermediate state (see text).}
\end{figure*}

\begin{figure*}[htb]
\centering
\includegraphics[width=15.0cm]{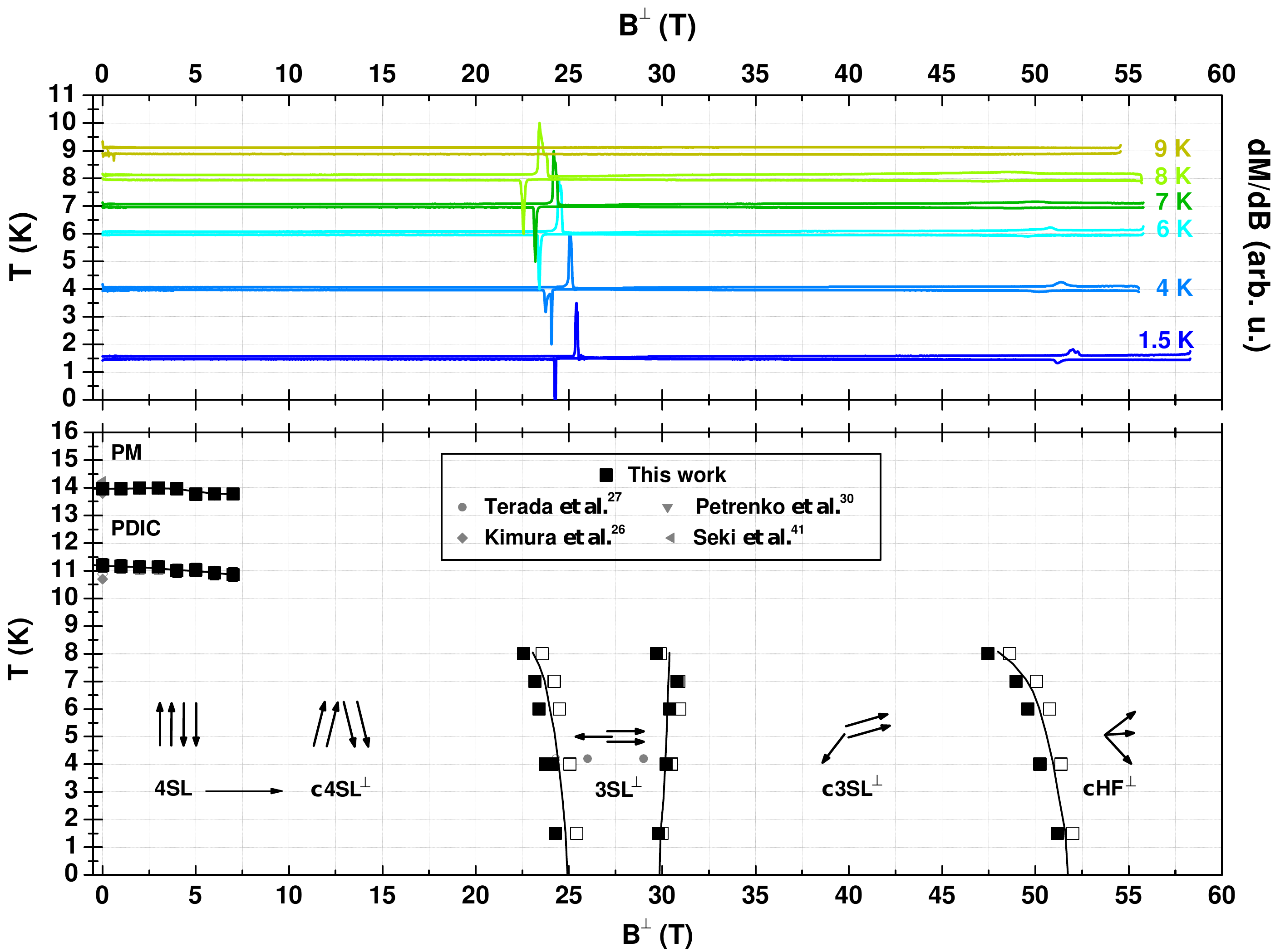}
\caption{\label{fig7} \small (Color online) Top panel: Differential
magnetization curves for $B\perp$ c. Colored solid and dotted lines
depict (normalized) $dM$/$dB$-curves as measured in increasing and
subsequently decreasing magnetic field, respectively.
$dM$/$dB$-curves measured in decreasing magnetic field are inverted
for clarity. Additionally, each curve has an offset equal to its
corresponding temperature for clarity. Bottom panel: $B$,$T$ phase
diagram of \CFO\ for the $B\perp$ c case. Large, black squares
depict magnetic transitions as observed in this work and smaller
gray symbols indicate previously reported transitions. Open and
closed symbols represent transitions observed in increasing and
decreasing ($B$ or $T$) ramps, respectively. Solid lines correspond
to proposed phase boundaries.}
\end{figure*}

Figure \ref{fig6} shows the $B$,$T$ phase diagram for \CFO\ that can
be constructed based on aforementioned experiments and other,
currently available literature
data\cite{petrenko00,petrenko05,kimura06,seki07,terada07_3,mitsuda00_1,mitsuda00_2,quirion09},
for the parallel configuration ($B\parallel$ c). The diagram
features all the previously confirmed phases; the zero field PM,
PDIC and 4SL phases and the consecutive FEIC $\rightarrow$ 5SL
$\rightarrow$ 3SL$^{\parallel}$ $\rightarrow$
\textit{c}3SL$^{\parallel}$ $\rightarrow$ \textit{c}HF$^{\parallel}$
phase cascade upon increasing field below \tnb. Worth noting is the
fact that the transition from the 5SL to the 3SL phase (at \Bparc)
appears to split up into a two-step transition with temperature,
implying an intermediate spin state $I$. At temperatures approaching
\tnb, the magnetization of the system in the corresponding field
region deviates continuously from the 3SL plateau value (see the 7 K
line in Figure \ref{fig2}), suggesting that here (some) spins are
canting away from collinearity, before the abrupt rearrangement to
the 5SL spin configuration. These double transition features were
observed before in steady state magnetic field measurements up to 23
T\cite{petrenko05}, which indicates that this behavior reflects the
inherent reduction of the magnetic anisotropy with applied magnetic
field in \CFO.\newline
\indent Figure \ref{fig7} shows the analogous $B$,$T$ phase diagram
resulting from above experiments and earlier reported
data\cite{petrenko00,kimura06,seki07,terada07_3} for the case where
$B\perp$ c. The diagram includes the zero field PM, PDIC and 4SL
phases and the field-induced phase cascade for this field
configuration: \textit{c}4SL $\rightarrow$ 3SL$^{\perp}$
$\rightarrow$ \textit{c}3SL$^{\perp}$ $\rightarrow$
\textit{c}HF$^{\perp}$.
\subsection{Classical spin model}
\begin{figure*}[htb]
\includegraphics[width=16.0cm]{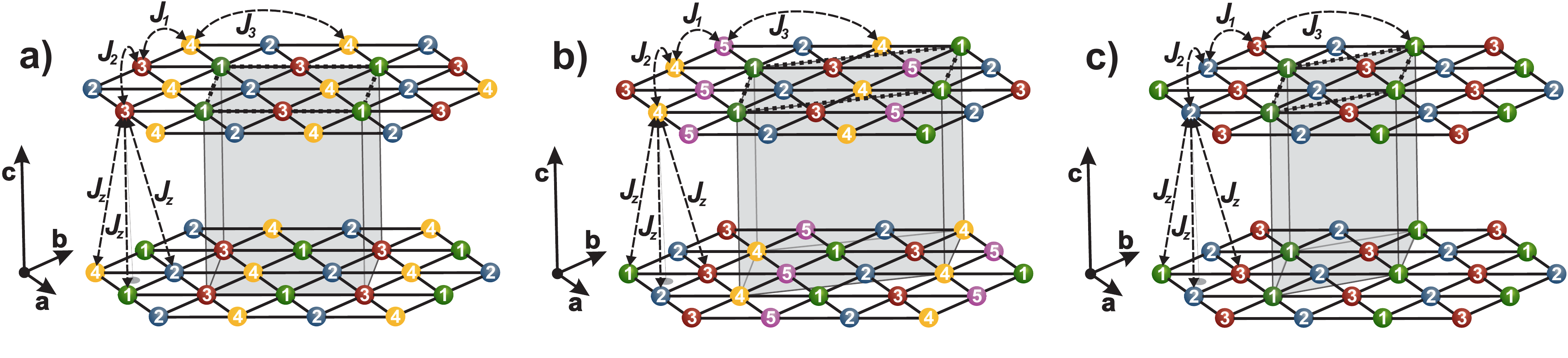}
\caption{\label{fig8} \small (Color online) Effective magnetic unit
cells in the consecutive \textbf{a)} four-, \textbf{b)} five- and
\textbf{c)} three-sublattice phases of \CFO. In all three structures
the dotted black lines outline the magnetic unit cell when a single
layer is considered (i.e. when $J_{z}=0$), while the grey shaded
volume indicates the effective two-layer magnetic unit cell
considered when interlayer interactions are taken into account.
Thus, an energetically optimal ABAB-type stacking is assumed for
each sublattice.}
\end{figure*}
\begin{figure*}[htb]
\includegraphics[width=15.0cm]{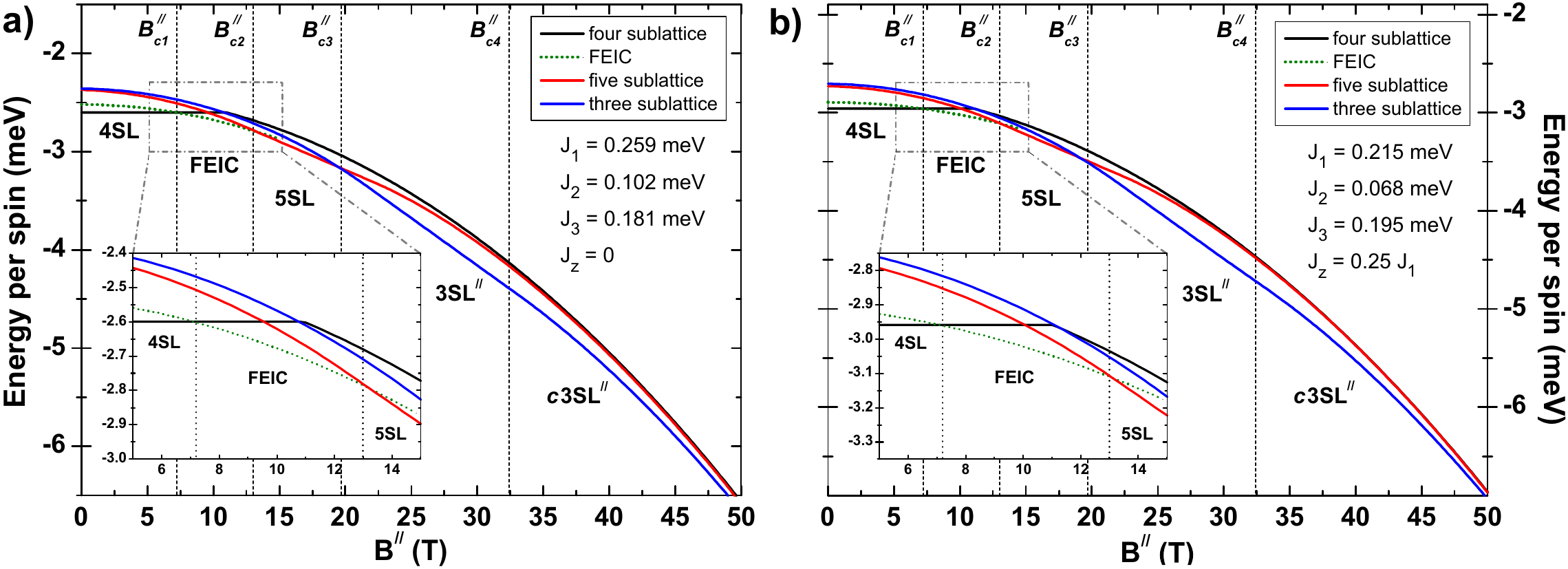}
\caption{\label{fig9} \small (Color online) Calculated minimum
energy per spin for each of the commensurate sublattice phases
($B\parallel$ c) as given by numerical minimization of equations
\ref{eq4}-\ref{eq6} without (\textbf{a)}) and with interlayer
exchange interactions (\textbf{b)}). Solid black, red and blue lines
correspond to the four-, five- and three-sublattice structures,
respectively. The dotted green line represents the expected energy
per spin of the incommensurate spiral (FEIC) phase. Dashed vertical
lines indicate experimental transition fields. Insets show a zoom-in
on the 5-15 T region, which features two magnetic phase
transitions.}
\end{figure*}
\begin{figure*}[htb]
\centering
\includegraphics[width=15.0cm]{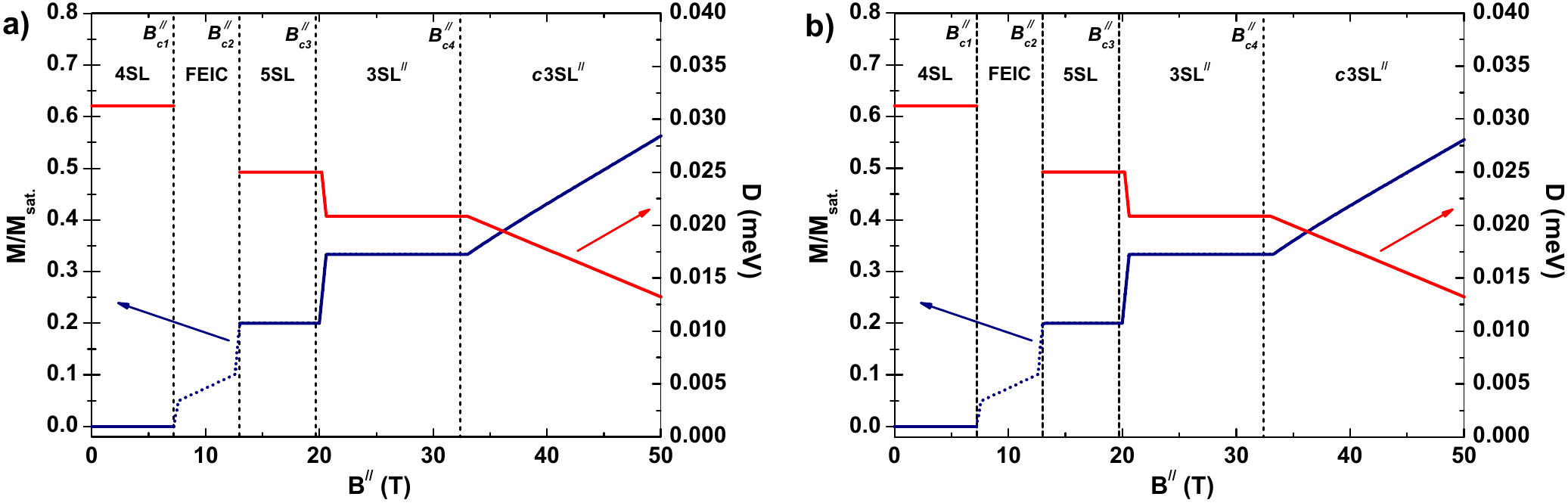}
\caption{\label{fig10} \small (Color online) Field dependence
($B\parallel$ c) of the magnetization and easy axis anisotropy in
\CFO\ for \textbf{a)} $J_{z}=0$ and \textbf{b)} $J_{z}=0.25J_{1}$.
Solid blue line: simulated magnetization process of \CFO\ at low
magnetic fields (see text). The blue dotted line in the FEIC phase
corresponds to the experimental data at 1.5 K (increasing field).
Solid red line: corresponding assumed values of the magnetic easy
axis anisotropy in the simulation of the various magnetic phases;
$D$ is approximated to be proportional to $(M_{sat.}-M(B))$.}
\end{figure*}
\begin{figure*}[htb]
\centering
\includegraphics[width=15.0cm]{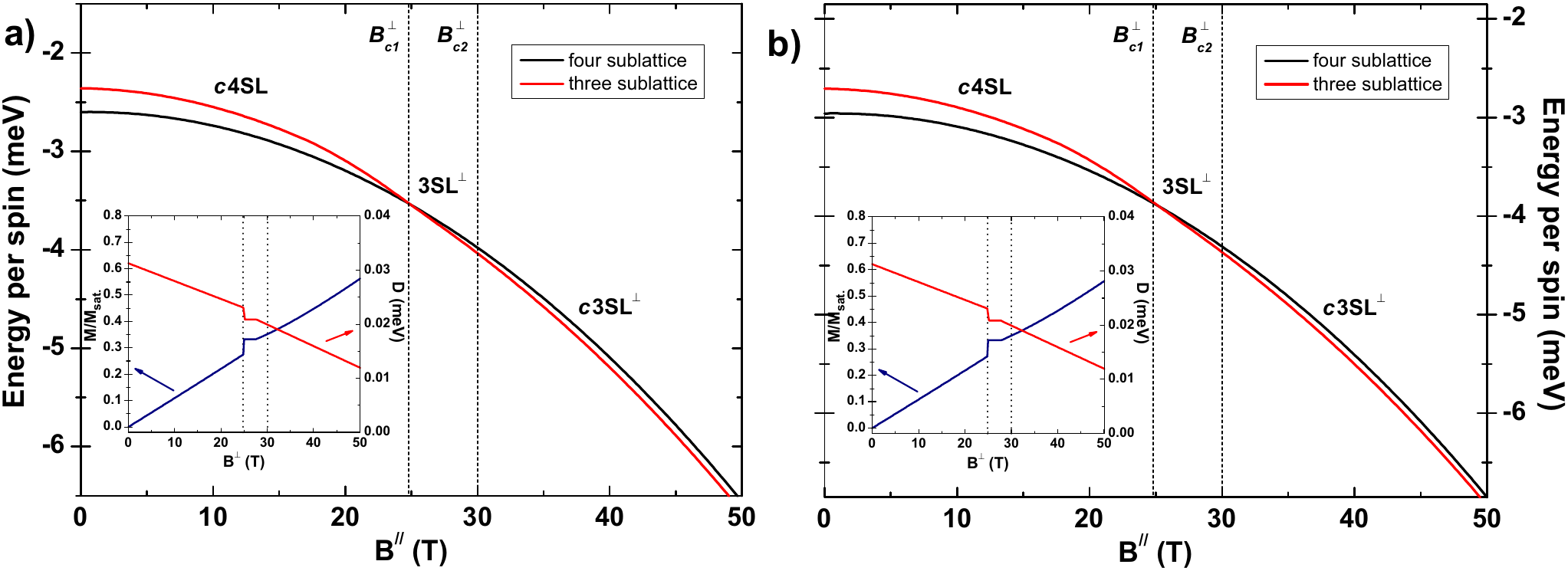}
\caption{\label{fig11} \small (Color online) Calculated minimum
energy per spin for the commensurate sublattice phases ($B\perp$ c)
for $J_{z}=0$ (\textbf{a)}) and $J_{z}=0.25J_{1}$ (\textbf{b)}).
Dashed vertical lines indicate experimental transition fields.
Inset: Corresponding field dependence of the simulated magnetization
(blue) and assumed easy axis anisotropy (red) in \CFO\ for $B\perp$
c.}
\end{figure*}

In order to study the magnetization process in \CFO\ further, we
resort to the PCS model. This phenomenological model includes the
primary magnetic interactions of the system; along with the basic
magnetic exchange and Zeeman interaction terms, the strong
spin-phonon coupling and the magnetic isotropy in \CFO\ are
included. The incorporation of the latter two seems key to capture
the Ising-like behavior of the system, as was recently
shown.\cite{wang08,lummen09} To determine the effect of interlayer
exchange interactions on the system, these are included in the model
separately later.\newline
\indent Spin-lattice interactions are typically incorporated into
the Hamiltonian through the distance dependence of the exchange
coupling $J(\textbf{r})$.\cite{penc04,bergman06,wang08} Ergo, in
general for a system with isotropic exchange interactions the
effective Hamiltonian becomes:
\begin{eqnarray}
H_{eff.} = J\displaystyle \sum_{\langle ij \rangle}
\mbox{\boldmath{$S$}}_{i}\cdot \mbox{\boldmath{$S$}}_{j}(1-\alpha
u_{ij}) + H_{def.}(\{\mbox{\boldmath{$u$}}_{i}\}),\label{eq1}
\end{eqnarray}
where the $\mbox{\boldmath{$u$}}_{i}$ are the displacement vectors,
the $u_{ij}$
(=$(\mbox{\boldmath{$u$}}_{i}-\mbox{\boldmath{$u$}}_{j})$$\cdot$$\mbox{\boldmath{$r$}}_{ij}$/$|r_{ij}|$)
are the corresponding relative changes in bond length between sites
$i$ and $j$, $\alpha$ is the spin-lattice constant (to first
approximation equal to $J^{-1}$ $\partial$$J$/$\partial$$r$) and
$H_{def.}$ corresponds to the deformation energy cost associated
with the atom displacements $\mbox{\boldmath{$u$}}_{i}$, which is
thus dependent on the phonon model still to be chosen. Taking the
simple bond-phonon (BP) model here, which treats the bond lengths
$u_{ij}$ as independent variables, the presence of spin-phonon
coupling effectively introduces an additional biquadratic spin
interaction of strength $bJ$, where $b$ = $\alpha^{2}J/k$ (third
term in eq. \ref{eq3}).\cite{penc04,bergman06} Furthermore, since
neighboring bond lengths $u_{ij}$ are independent here, the
biquadratic term is restricted to nearest neighbor couplings only.
Due to the quadratic nature of the term, either parallel or
antiparallel spin configurations are favorable, which explains the
tendency of spin-lattice coupling to stabilize collinear spin
states.\newline \indent Thus, the general spin Hamiltonian
(containing only magnetic contributions) for \CFO\ within the PCS
model\cite{lummen09} can now be constructed:
\begin{eqnarray}
H_{s} = & -g\mu \mbox{\boldmath{$B$}}\cdot\displaystyle
\sum_{i}\mbox{\boldmath{$S$}}_{i} + \displaystyle \sum_{i,j}
J_{ij}\mbox{\boldmath{$S$}}_{i}\cdot
\mbox{\boldmath{$S$}}_{j} \nonumber\\
&  - \displaystyle \sum_{\langle ij \rangle}
bJ_{ij}(\mbox{\boldmath{$S$}}_{i}\cdot
\mbox{\boldmath{$S$}}_{j})^{2} -
D(\mbox{\boldmath{$B$}})\displaystyle \sum_{i} S_{iz}^2,\label{eq3}
\end{eqnarray}
where $\mbox{\boldmath{$B$}}$ is the applied magnetic field,
$J_{ij}$ is the exchange interaction between sites $i$ and $j$, $b$
is the biquadratic coupling constant and $D$ is the magnetic
anisotropy constant, which is field dependent due to its strong
coupling to the lattice distortion. The Zeeman and anisotropy terms sum over all sites
$i$, the biquadratic term couples only nearest neighbor spin pairs
$i$ and $j$, and the exchange term includes all spin pair
interactions in the system.\newline
\indent In a previous work, we analyzed the behavior of this spin
Hamiltonian (eq. \ref{eq3}) when applied to the magnetic unit cell
of the three sublattice structure, thereby focusing on the high
field magnetic phases of \CFO. Here, we compare the field dependence
of all consecutive commensurate phases, based on the same spin
Hamiltonian and the previously extracted parameters. Thus, we
evaluate the corresponding spin Hamiltonians for the magnetic unit
cells of the four-, five- and three-sublattice structures on a
single triangular sheet; the corresponding unit cells are sketched
in Figure \ref{fig8}. Considering the spins as classical, justified
by the large $S = 5/2$ value, we write
$\mbox{\boldmath{$S$}}_{i}=\mbox{\boldmath{$e$}}_{i}S$ (where
$\mbox{\boldmath{$e$}}$ is a unit vector), and $J_{1}$, $J_{2}$ and
$J_{3}$ for the first, second and third nearest neighbor exchange
interactions, respectively. The respective spin Hamiltonians are
then found to be:

\begin{eqnarray}
\lefteqn{H_{4SL} = }\nonumber \\
&&-2\mu_{B}S\mbox{\boldmath{$B$}}\cdot\displaystyle
\sum_{i}\mbox{\boldmath{$e$}}_{i}-{}D(B)S^{2}\displaystyle
\sum_{i} e_{i,z}^2 \nonumber \\
&& { }+{}2J_{1}S^{2}(p_{12}+p_{13}+p_{14}+p_{23}+p_{24}+p_{34}) \nonumber \\
&& { }{ }+{}2J_{2}S^{2}(2+p_{12}+p_{14}+p_{23}+p_{34}) \nonumber \\
&& { }{ }{ }+{}4J_{3}S^{2}(1+p_{13}+p_{24}) \nonumber \\
&&
{}{}{}{}-{}2AS^{4}(p_{12}^{2}+p_{13}^{2}+p_{14}^{2}+p_{23}^{2}+p_{24}^{2}+p_{34}^{2}),\label{eq4}
\end{eqnarray}
\begin{eqnarray}
\lefteqn{H_{5SL} = }\nonumber \\
&&-2\mu_{B}S\mbox{\boldmath{$B$}}\cdot\displaystyle
\sum_{i}\mbox{\boldmath{$e$}}_{i}-{}D(B)S^{2}\displaystyle
\sum_{i} e_{i,z}^2 \nonumber \\
&& +{}2J_{1}S^{2}(p_{12}+p_{15}+p_{23}+p_{34}+p_{45}) \nonumber \\
&& +{}J_{1}S^{2}(p_{13}+p_{14}+p_{24}+p_{25}+p_{35}) \nonumber \\
&& +{}J_{2}S^{2}(5+2p_{13}+2p_{14}+2p_{24}+2p_{25}+2p_{35}) \nonumber \\
&& +{}2J_{3}S^{2}(p_{13}+p_{14}+p_{24}+p_{25}+p_{35}) \nonumber \\
&& +{}J_{3}S^{2}(p_{12}+p_{15}+p_{23}+p_{34}+p_{45}) \nonumber \\
&&-{}2AS^{4}(p_{12}^{2}+p_{15}^{2}+p_{23}^{2}+p_{34}^{2}+p_{45}^{2}) \nonumber \\
&&-{}AS^{4}(p_{13}^{2}+p_{14}^{2}+p_{24}^{2}+p_{25}^{2}+p_{35}^{2}),\label{eq5}
\end{eqnarray}
\begin{eqnarray}
\lefteqn{H_{3SL} = }\nonumber \\
&&-2\mu_{B} S\mbox{\boldmath{$B$}}\cdot\displaystyle
\sum_{i}\mbox{\boldmath{$e$}}_{i}-{}D(B)S^{2}\displaystyle \sum_{i}
e_{i,z}^2 \nonumber \\
&& +{}3(J_{1}+J_{3})S^{2}(p_{12}+p_{13}+p_{23})\nonumber \\
&& +{}9J_{2}S^{2} \nonumber \\
&& -{}3AS^{4}(p_{12}^{2}+p_{13}^{2}+p_{23}^{2}),\label{eq6}
\end{eqnarray}
where $g$ is taken as 2 and spin-spin couplings are written as
$p_{ij}$
$(=\mbox{\boldmath{$e$}}_{i}\cdot\mbox{\boldmath{$e$}}_{j}$). The
exchange constants are taken as equal along the different in-plane
crystallographic directions, their field-dependence being in the
spin-phonon term. The spin-phonon parameter is defined as $A=bJ_{1}$, which
corresponds to $G/3$ in our previous work\cite{lummen09}, though
with a rescaled dimensionless biquadratic coupling $b$ of $\simeq$
0.0098 (here, third nearest neighbour interactions are taken into
account in the estimation of $b$).\newline
\indent To test the PCS model, we performed numerical minimization
of equations \ref{eq4}-\ref{eq6} as a function of the independent
spin vectors ($\mbox{\boldmath{$e$}}_{1}$,
$\mbox{\boldmath{$e$}}_{2}$, $\mbox{\boldmath{$e$}}_{3}$,
$\mbox{\boldmath{$e$}}_{4}$ and $\mbox{\boldmath{$e$}}_{5}$) at a
given field $B$, using previously extracted parameters. In order to
incorporate its field dependence, which is \textit{a priori}
unknown, $D$ is approximated to be proportional to $(M_{sat.}-M(B))$
here. In other words, $D$ is assumed to mirror the field-dependence
of $M$, undergoing stepwise reductions at first order transitions
and vanishing as the system approaches saturation; see Figures
\ref{fig10} and \ref{fig11}. The previously estimated $D$ for the
collinear 3SL phases (3SL$^{\parallel}$ and 3SL$^{\perp}$) was
$\simeq$ 0.021 meV, making it $\simeq$ 0.031 meV and $\simeq$ 0.025
meV in the collinear 4SL and 5SL phases, respectively. Taking
exchange couplings as $J_{1}$ $\simeq$ 0.259 meV, $J_{2}$ $\simeq$
0.102 meV and $J_{3}$ $\simeq$ 0.181 meV, and the spin-phonon
parameter $A$ as $\simeq$ 0.00247 meV, the parallel-field dependence
($B\parallel$ c) of the resulting energy per spin for each of the
commensurate sublattice phases is as shown in Figure
\ref{fig9}a.\newline
\indent Upon examination of the different energy curves, one finds
that the PCS model with these parameters yields a cascade of
expected magnetic transitions that is consistent with experiment.
The 4SL collinear four-sublattice state is stable with respect to
the 5SL structure up to $\simeq$ 9.4 T. From there on, the 5SL state
is the most energetically favorable, up to the critical field
\Bparc. Above \Bparc, the collinear 3SL$^{\parallel}$ state becomes
stable, undergoing a transition to the $c$3SL$^{\parallel}$
structure only around \Bpard\ $\simeq$ 32.4 T. Experimentally, the
multiferroic spiral FEIC phase was found as an intermediate phase,
between \Bpara\ $\simeq$ 7.2 T and \Bparb\ $\simeq$ 13.0 T. As this
phase is incommensurate, however, it is not feasible to describe it
using the PCS model applied to a limited-size unit cell here.
Recently though, such complex incommensurate ground state structures
were found in zero field for far larger unit cells using Monte Carlo
simulations.\cite{fishman10,haraldsen09} Based on the experimental
data, the energy per spin of the field-induced FEIC phase in \CFO\
is expected to have a field-dependence as indicated by the green
dotted line in Figure \ref{fig9}a, making it the adopted spin
structure between \Bparb\ and \Bparc.\newline
\indent The corresponding magnetization curve for $B\parallel$ c,
depicted in Figure \ref{fig10}a, shows a good agreement with the
experimental result (Figure \ref{fig2}, 1.5 K curve). The
non-directional spin-phonon interaction ($A$), which favors
collinear spin states, combines with the directional applied field
$B^{\parallel}$ and the easy axis anisotropy $D(B)$ to successively
stabilize the consecutive magnetization plateaus of the collinear
phases. At high fields (above \Bpard), the increasingly dominant
Zeeman term and the progressively reduced anisotropy result in a
gradual spin canting in the system.\newline
\indent In analogous fashion, one can calculate the energy per spin
for the commensurate sublattice phases in case of a field applied
perpendicular to the c-axis using equations \ref{eq4} and \ref{eq6},
respectively. Using the same parameters as used for the $B\parallel$
c case, one obtains an energy scheme as depicted in Figure
\ref{fig11}a. The $c$4SL state is the most energetically favored up
to \Bperpa\ $\simeq$ 24.8 T, above which a three-sublattice is the
most stable, with the spins adopting consecutive 3SL$^{\perp}$ and
3SL$^{\perp}$ structures as the applied field increases. The inset
of Fig. \ref{fig11}a shows the corresponding simulated magnetization
curve for $B\perp$ c, as well as the corresponding assumed value of
the magnetic easy axis anisotropy in the various magnetic phases.
The obtained magnetization process is again in good agreement with
the experimental curve (Figure \ref{fig4}, 1.5 K line). As opposed
to the $B\parallel$ c case, the directional anisotropy is orthogonal
to the field direction here, resulting in a much smaller plateau
width. Thus, the PCS spin Hamiltonian (eq. \ref{eq3}) also provides
an adequate description of the low field part of the magnetization
process in \CFO, for both field configurations, using the same
parameters that were previously used for describing the high field
part.\newline
\indent We emphasize the fact that the spin Hamiltonian parameters
used were determined through direct comparison with experimentally
observed features. The easy axis anisotropy $D$ (only a scaling
parameter as $D(B)$ $\propto$ $(M_{sat.}-M(B))$) and spin-phonon
coupling $A$ were determined through the simulation of the
high-field magnetization process, which also set the value for the
sum of $J_{1}$ and $J_{3}$.\cite{lummen09} With these preset
restrictions, $J_{2}$ and $J_{3}$ were set such that: \textit{i.}
the simulated 5SL to 3SL$^{\parallel}$ transition field for
$B\parallel$ c corresponds to the experimental value (\Bparc), and
\textit{ii.} the simulated \textit{c}4SL to 3SL$^{\perp}$ transition
field corresponds to the experimental \Bperpa\ value. The resulting
exchange parameters compare as $J_{2}/J_{1}$ $\simeq$ 0.39 and
$J_{3}/J_{1}$ $\simeq$ 0.70, ratios which are close to those
previously estimated.\cite{ye07,fishman08}\newline
\subsection{Interlayer exchange interaction}
Recent efforts suggest a magnetic exchange interaction between the
Fe-layers to be an additional important aspect of the \CFO\ system.
Inelastic neutron scattering work shows indicative spin-wave
dispersion along the hexagonal axis, signaling the interlayer
interaction to be
significant.\cite{petrenko05,terada07_2,ye07,fishman08} This is
corroborated by the observation of finite dispersion of calculated
electronic bands.\cite{eyert08} Thus, here we incorporate the
interlayer exchange into the PCS model to determine its influence on
the modeled magnetization process. As the interplane interaction is
estimated to be small compared to the in-plane exchange, we take a
perturbative approach, taking only nearest neighbor interactions
($J_{z}$\textgreater0). With each spin having nearest neighbor
interlayer couplings to three consecutive sublattice sites in the
adjacent layers, all types of stacking of three-sublattice layers
are energetically equivalent, while four- and five-sublattice layers
have specific optimum stacking sequences (those depicted in figure
\ref{fig8}).\cite{fishman08,note} Assuming this optimal stacking of
consecutive layers, the effective magnetic unit cells of the four-,
five- and three-sublattice structures now contain two triangular
sheets each, with the additional interlayer interactions amounting
to:
\begin{eqnarray}
H_{z,4SL}&=&2J_{z}S^{2}(p_{12}+p_{13}+p_{14}+p_{23}+p_{24}+p_{34}) \\
H_{z,5SL}&=&2J_{z}S^{2}(p_{13}+p_{14}+p_{24}+p_{25}+p_{35})\nonumber \\
&&+{}J_{z}S^{2}(p_{12}+p_{15}+p_{23}+p_{34}+p_{45}) \\
H_{z,3SL}&=&J_{z}S^{2}(3+2p_{12}+2p_{13}+2p_{23})
\end{eqnarray}
per layer. For the collinear sublattice structures these terms add
up to $-J_{z}S^{2}$, $-J_{z}S^{2}$ and $+J_{z}S^{2}/3$ per spin for
the four-, five- and three-sublattice structures, respectively. With
the inclusion of these terms, equations \ref{eq4}-\ref{eq6} were
once again numerically minimized to determine optimum spin
directions in an increasing field; the resulting energies of the
different sublattices and the corresponding simulated magnetization
curves are depicted in Figures \ref{fig9}b and \ref{fig10}b for
$B\parallel$ c and in Figure \ref{fig11}b for $B\perp$ c. As is
clear from these graphs and their comparison to the case where
$J_{z} = 0$, the experimental magnetization process is equally well
simulated upon incorporation of interlayer interactions. Keeping $A$
and $D(B)$ at the same value, the incorporation of $J_{z}$, which
was fixed at 0.25$J_{1}$ (a representative value based on inelastic
neutron scattering data\cite{ye07,fishman08}), results in adapted
extracted exchange couplings of $J_{1}$ $\simeq$ 0.215 meV (making
$J_{z}$ $\simeq$ 0.054 meV), $J_{2}$ $\simeq$ 0.068 meV and $J_{3}$
$\simeq$ 0.195 meV. As before, these parameters were determined
through direct comparison with observed experimental features of the
magnetization process. Though the introduction of an additional
antiferromagnetic interaction in the model generally tends to
decrease the extracted parameters, $J_{3}$ is in fact increased here
to counter the relative destabilization of the three-sublattice
structure. Summarizing, incorporation of interlayer interactions
into the PCS model yields an equally adequate description of the
experimental magnetization process of \CFO, with slightly modified
exchange parameters.\newline
\indent At this point, it is worth pointing out the limitations of
the PCS model presented here. As our calculations focus on
minimizing the magnetic energy in specific, chosen sublattice
structures, other possible commensurate or incommensurate states are
effectively neglected. Calculations on larger magnetic unit cells or
triangular lattices with periodic boundary conditions may uncover
larger sublattice or more complex spin configurations within the
model that may be relevant, as was found to be the case for the
zero-field phase of doped \CFO.\cite{haraldsen09,fishman10} The
recently proposed incommensurate 120\ensuremath{^\circ}-like spin
structure above \Bpare\ is one example, though its underlying Landau
theory does not capture some general features of the experimental
high-field magnetization curve at present.\cite{quirion09} A fully
accurate and quantitative description of the magnetism of \CFO\
would require the inclusion of all additional features of the system
that could play a role. The incorporation of finite temperature, a
more realistic phonon model (yielding longer range biquadratic
interactions\cite{wang08,bergman06}) and quantum spin effects may
improve the quantitative understanding of the system. Furthermore,
more exotic interactions may play a role in stabilizing the
incommensurate spiral state.\cite{plumer08} Nevertheless, the simple
PCS model presented here is shown to capture almost all general
features of the experimental magnetization process in both field
configurations, providing a satisfactory and intuitive description
of the observed magnetism in \CFO.
\section{Conclusions}
Summarizing, we have performed magnetization experiments on \CFO\ at
various temperatures below \tnb\ up to high magnetic fields, both
for $B\parallel$ c and $B\perp$ c field configurations. The
characteristic magnetic staircase of \CFO\ was reproduced and found
to retain its general features with increasing temperature below
\tnb. As the temperature approaches \tnb\ however, transition
features are progressively smoothed out and plateau phases are found
to acquire increasing slopes. Moreover, the transition from the
collinear 5SL to the collinear 3SL$^{\parallel}$ phase (at \Bparc)
was shown to split up into a two-step transition near \tnb,
revealing an additional, possibly noncollinear, intermediate state
$I$ at these temperatures. Additionally, the various critical fields
of the same nature are shown to exhibit a very similar temperature
dependence; all first order transitions exhibit an analogous
relative decrease with temperature, and second order transitions are
found to be relatively temperature independent. Correspondingly, we
have thoroughly mapped out the experimental $B$,$T$ phase diagrams
of \CFO\ for both the parallel ($B\parallel$ c) and perpendicular
($B\perp$ c) configurations and expanded them in both temperature
and magnetic field. Through numerical minimization of the PCS model
applied to the consecutive commensurate sublattice phases of \CFO,
also the low-field part of the experimental magnetization process
was adequately simulated, yielding reasonable estimates for the
additional parameters $J_{2}$ and $J_{3}$. Incorporation of an
additional interlayer exchange interaction in the model was shown to
result in a nearly identical simulation and a somewhat adapted set
of exchange interactions. Thus, the proposed PCS model, combined
with the underlying notion of progressive symmetry increase with
applied field, is found to provide a satisfactory semi-quantitative
description of the entire magnetization process of \CFO.\newline

\begin{acknowledgments}
This work is dedicated to the memory of Harison Rakoto. The authors
would like to thank A.A. Nugroho for his help in the single crystal
growth, T.T.M. Palstra and J. Baas for facilitating use of the
optical floating zone furnace and SQUID and F. de Haan and D.
Maillard for technical support. Financial support from the Agence
Nationale de Recherche under contract NT05-4\_42463 is gratefully
acknowledged.
\end{acknowledgments}


\begin{thebibliography}{47}
\bibitem[{not()}]{note}
\bibinfo{note}{The five-sublattice structure has other types of optimal
  stacking which are energetically equivalent.}
\end{thebibliography}
\end{document}